\documentstyle[amssymb,epsfig,aps]{revtex}

\begin{document}
\twocolumn

\title{Squeezing of Atoms in a Pulsed Optical Lattice}
\author{M. Leibscher and I. Sh. Averbukh}
\address{Department of Chemical Physics, The Weizmann Institute of Science, Rehovot\\
76100, Israel\\
e-mail: Ilya.Averbukh@weizmann.ac.il}
\maketitle

\begin{abstract}
We study the process of squeezing of an ensemble of cold atoms in a pulsed
optical lattice. The problem is treated both classically and
quantum-mechanically under various thermal conditions. We show that a
dramatic compression of the atomic density near the minima of the optical
potential can be achieved with a proper pulsing of the lattice. Several
strategies leading to the enhanced atomic squeezing are suggested, compared
and optimized. 

\end{abstract}

\section{Introduction}

Optical lattices are periodic potentials for neutral atoms induced by
standing light waves formed by counter-propagating laser beams. When these
waves are detuned from any atomic resonance, the ac Stark shift of the
ground atomic state leads to a conservative periodic potential with spatial
period $\lambda /2,$ half the laser wavelength (for a review, see, e.g. \cite
{Jessen}). Such lattices present a convenient model system for solid state
physics and nonlinear dynamics studies. In contrast to traditional solid
state objects, the parameters of optical lattices (i.e. lattice constant,
potential well depth, etc.) are easily controllable. Many fine phenomena
that were long discussed in solid state physics, have been recently observed
in corresponding atom optics systems. Bloch oscillations \cite{Dahan}, or
the Wannier-Stark ladder \cite{Wilkinson,Niu} are only few examples to
mention. Time-modulation of the frequency and intensity of the constituent
laser beams provide tools for effective modeling of numerous time-dependent
nonlinear phenomena. Since the initial proposal \cite{Graham}, and first
pioneering experiments on atom optics realization of the delta-kicked
quantum rotor \cite{Moore,Ammann}, cold atoms in optical lattices provide
also new grounds for experiments on quantum chaos. Optical lattices are
directly related to the technologically important problem of atom
lithography. Potential wells of a standing light wave may serve as a
periodic array of focusing elements for an atomic beam in a deposition
setup. Atom lithographic schemes based on nanomanipulation of neutral atoms
by laser fields have attracted a lot of attention recently \cite
{Timp,McClelland,Cellota,Drodofsky,Gupta,review}, as a promising way to
overcome the resolution limits of usual optical lithography. In the case of
a direct laser-guided atom deposition, the diffraction resolution limit is
determined by the de Broglie wavelength of atoms, and may reach several pm
for typical atomic beams. In practice, however, this limit has never been
relevant, mainly because of severe aberration of the sinusoidal potential of
a standing light wave. As a result, all current atom lithography schemes
suffer from a considerable background in the deposited structures. A
possible way to overcome the problem was suggested in \cite{Schleich}, by
using nanofabricated mechanical masks that block atoms passing far from the
potential minima of the optical lattice. However, this complicates
considerably the setup and reduces the deposition rate. Therefore, there is
a considerable need in pure atomic optics solutions of the aberration
problem. In paraxial approximation, the steady-state propagation of an
atomic beam through a standing light wave is closely connected to the
problem of the time-dependent lateral motion of atoms subject to a spatially
periodic potential of an optical lattice. The focusing phenomenon
corresponds to the narrowest state reached in the course of a ''breathing''
motion of the atomic wave packet. From this point of view, an enhanced
focusing may be considered as a kind of {\em squeezing } {\em problem}
well-known in quantum optics. The established way to induce squeezing in a
harmonic system is through parametric modulation of its spring constant. In
the case of rather cold atoms and/or deep optical lattices, in which the
atomic groups are localized near the potential minima, the harmonic
approximation works reasonably well. The parametric excitation needed for
squeezing may be achieved by modulating the intensity of the laser beams. A
transient compression of atoms in optical lattices was demonstrated via this
mechanism \cite{Gorlitz,Raithel1,Bigelow,Monroe}, using Bragg scattering 
\cite{Birkl,Hansch,Raithel2}, or time-resolved fluorescence \cite{Bigelow}
as a sensitive probe for the atomic localization. Manipulation of
the breathing modes of the atomic oscillations in an optical trap was
considered in \cite{Burin97,Bulatov98,Bulatov99} as a tool for optical cooling 
and controlling the onset of Bose- Einstein condensation. 

We note, however, that  no considerable squeezing can be
achieved for any periodic parametric driving of the sinusoidal potential
because of its anharmonicity. 
This problem happens to be closely related to the physics of orienting
(aligning) molecules by strong laser fields \ (see, e.g. \cite
{Normand,Dietrich,Friedrich,Corkum,Tamar1,Tamar2,Cai}, and references
therein). In recent work \cite{rotor}, generic properties of a strongly\
driven quantum rotor were analyzed, and its dynamics was shown to be subject
to a number of spectacular semiclassical catastrophes. The transient
orientation of the rotor by time-dependent fields was considered as a
nonlinear squeezing problem, and a strategy was demonstrated based on
aperiodic driving of the rotor by a specially designed sequence of short
laser pulses, which allows for accumulative compression of the rotor angular
distribution. Some related squeezing approaches were recently suggested as
efficient tools for atom lithography of ultra-high resolution \cite{patent}.

In the present paper we study in detail the squeezing aspects of atomic
behavior in pulsed optical lattices. In Sec. II the formal framework of the
problem is defined and the squeezing of a classical thermal ensemble of
atoms is studied. Here we examine the accumulative squeezing scheme of \cite
{rotor} in application to atom optics systems, and also consider some more
refined optimization approaches. In Sec. III the problem is treated in the
quantum-mechanical domain, in which some new aspects appear at low
temperatures. Finally, the results of the paper are summarized in the
concluding Sec. IV.

\section{Squeezing of atoms: classical treatment}

We consider a system of cold atoms interacting with a pulsed standing wave
of light. The atoms are described as two level systems with transition
frequency $\omega _{0}$. The standing wave is linearly polarized and has a
frequency of $\omega _{l}$. If the detuning $\Delta _{l}=\omega _{0}-\omega
_{l}$ is large compared to the relaxation rate of the excited atomic state,
its population can be adiabatically eliminated \cite{Graham}. In this case,
the internal structure of the atoms can be completely neglected and they can
be regarded as point-like particles. In this approximation, the Hamiltonian
for the atomic motion across the standing wave is 
\begin{equation}
H(x,p)=\frac{p^{2}}{2m}-V(t)\cos \left( 2k_{l}x\right) .  \label{Hamiltonian}
\end{equation}
Here $m$ is the atomic mass, and $k_{l}=\omega _{l}/c$ is the wave number of
laser beams forming the standing wave. The depth of the potential produced
by the standing wave is $V(t)=\hbar \Omega (t)^{2}/(8\Delta _{l})$, where $%
\Omega (t)=2d_{z}E(t)/\hbar $ is the Rabi frequency, $\vec{d}$ is the atomic
dipole moment, and $E(t)$ is the slowly varying amplitude of the light
field. We consider the case of a red-detuned optical lattice ($\Delta _{l}>0$%
), i.e. high-field seeking atoms.

Many aspects of the dynamics of atoms in a pulsed optical lattice which will
be discussed below, can be explained with semiclassical arguments.
Therefore, we start with considering the problem in the classical framework.
If the initial atomic distribution is spatially uniform, it will become
periodic in space (with the optical lattice period) for any time dependence
of \ $V(t)$. Under these conditions, the dynamics of our system is similar
to the dynamics of an ensemble of two-dimensional rotors, or, equivalently,
of an ensemble of particles freely moving on a ring. We are especially
interested in the case of rather short laser pulses, when Hamiltonian Eq.(%
\ref{Hamiltonian}) corresponds to that of the $\delta $-kicked rotor. Using
dimensionless variables, we can express the atom velocity after a short kick
provided by the pulsed optical lattice as 
\begin{equation}
v^{\prime }=v_{0}^{\prime }-\sin x_{0}^{\prime },  \label{clv}
\end{equation}
\begin{figure}[h]
  \center{\epsfig{file=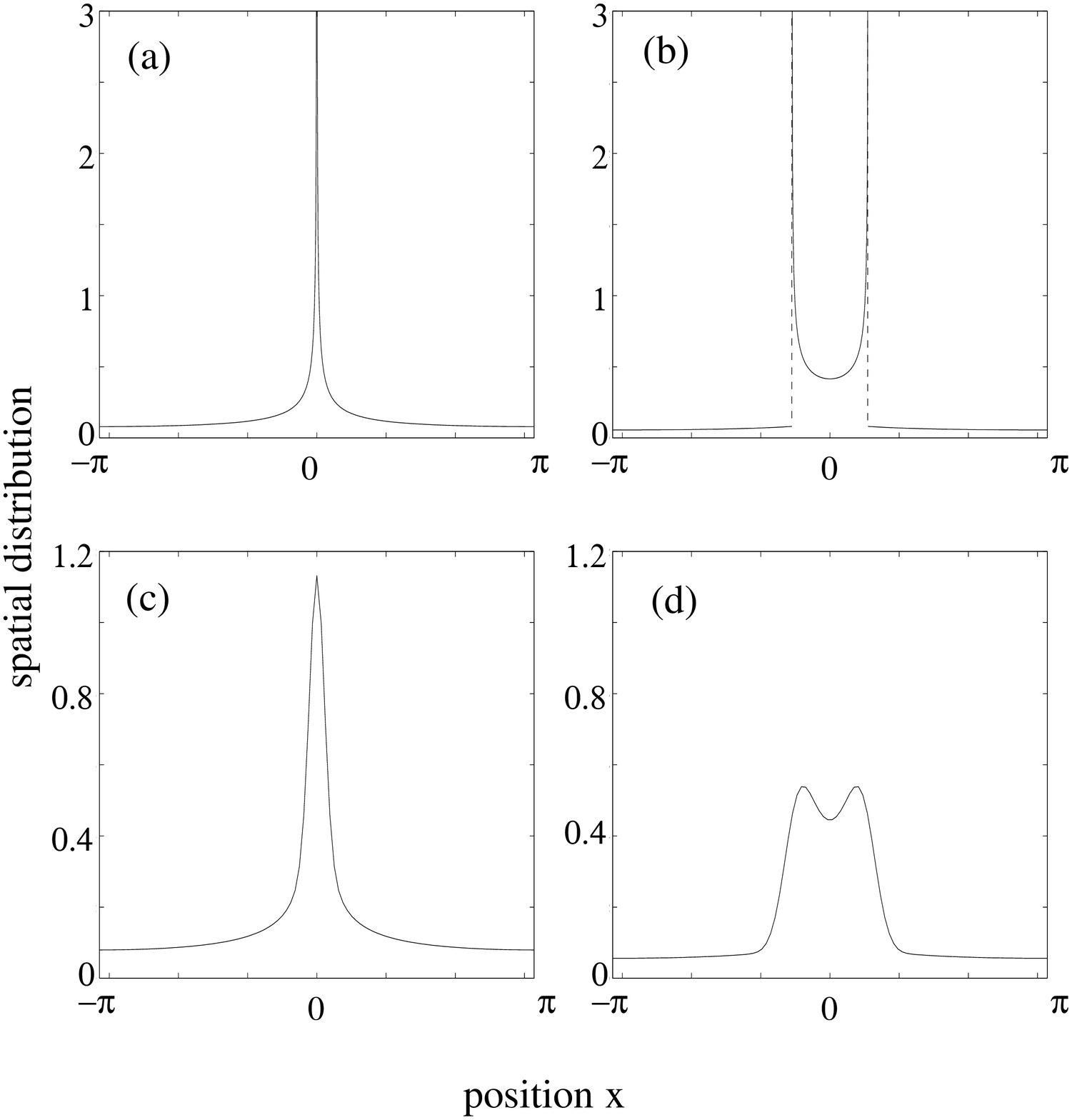,width=8.6 cm, angle=0}}
  \caption{Spatial distribution for a classical ensemble of atoms after a
single $\delta$-kick. In Figs. (a) and (b), the initial temperature T of the
ensemble is zero. In (c) and (d), the width of the thermal initial
distribution is $\sigma_{cl}=0.1$. Figures (a) and (c) show the spatial
distribution at focal time $\tau=\tau_f$, while in Figs. (b) and (d) $%
\tau=1.84$.}
\end{figure}
Here $v_{0}^{\prime }=mv_{0}/[2k_{l}\int V(t)dt]$ is the dimensionless
velocity before the kick, and $x_{0}^{\prime }=2k_{l}x_{0}$ is the
(dimensionless) position of the atom at the kick time. The integration in $%
\int V(t)dt$ is done over the duration of the short pulse. Between the
pulses, the atom moves with a constant velocity, and its position evolves as 
\begin{equation}
x^{\prime }(\tau )=x_{0}^{\prime }+v^{\prime }\tau \ \ \ (\mbox{mod}\ \ 2\pi
).  \label{clx}
\end{equation}
Here, $\tau =4k_{l}^{2}\left[ \int V(t^{\prime })dt^{\prime }\right] \,\,t/m$
is the dimensionless time. (In the following, we drop the primes in Eqs. (%
\ref{clv}) and (\ref{clx}) but keep in mind that we are using the
dimensionless variables).

The probability of finding an atom in a certain phase-space element $dx\,dv$
at time $\tau $ after a kick is then determined by the initial distribution
function $f(x_{0},v_{0},\tau =0)$ as follows 
\begin{equation}
f(x,v,\tau )dx\,dv=f\left[ x-vt,v+\sin (x-v\tau ),\tau =0\right] dxdv.
\end{equation}
For a thermal initial velocity distribution, the probability density at time 
$\tau $ is then 
\begin{equation}
f(x,v,\tau )=\frac{1}{(2\pi )^{3/2}\sigma _{cl}}\exp \left\{ -\frac{\left[
v+\sin \left( x-v\tau \right) \right] ^{2}}{2\sigma _{cl}^{2}}\right\} ,
\label{disfun}
\end{equation}
where $\sigma _{cl}$ is associated with the temperature, $T$ via 
\begin{equation}
\sigma _{cl}=\frac{\sqrt{k_{B}Tm}}{2k_{l}\int_{-\infty }^{\infty }V(t)dt}.
\label{sigma}
\end{equation}
Here, $k_{B}$ is the Boltzmann constant. By integrating Eq.(\ref{disfun})
over $v$, we obtain the spatial distribution function $f(x,\tau)$. For zero
initial temperature, the Gaussian velocity distribution in Eq.(\ref{disfun})
becomes a $\delta $-function, and the spatial distribution function is 
\begin{equation}
f(x,\tau )=\frac{1}{2\pi }\sum_{a}\frac{1}{\left| dx/dx_{0}^{a}\right| }.
\end{equation}
Here the summation is performed over all possible branches of the function $%
x_{0}(x)$ defined by Eqs. (\ref{clv}) and (\ref{clx}), as described in \cite
{rotor}. Figure 1 displays the spatial distribution of atoms at different
times after a single $\delta $-kick. In pictures (a) and (b), the initial
temperature of the ensemble is zero. Figure 1 (a) shows the spatial
distribution at $\tau =1$ after the pulse. One observes a sharp, narrow peak
at $x=0$ above a broad background. The physics behind this feature is
similar to the focusing of light rays by an optical lens. Indeed, the
velocity of an atom being initially at the position $x$ is $v=-\sin x$.
Therefore, atoms with $x<<1$ have a velocity proportional to their initial
position, and they all arrive at the focal point $x=0$ at the same focusing
time, $\tau _{f}=1$. In this language, the finite width of the focal spot
and the broad background are due to the effect of ''spherical aberrations'',
that means, due to the deviation of the $\cos x$ potential from the
parabolic one.\ Figures 1 (b) shows the spatial distribution at $\tau
=1.84\tau _{f}$, the time of the maximal squeezing (see below). The
distribution is characterized by two singularities. The origin of these
peaks is due to the coexistence of two counter-propagating groups of atoms,
which appear near $x=0$ at $\tau >\tau _{f}$: those that were focused, and
those that were not fast enough to reach the focal point $x=0$ at $\tau
=\tau _{f}$. This effect is similar to the formation of rainbows in the wave
optics \cite{Kravtsov} and quantum mechanics \cite{Ford,Berry}, and it is
discussed in detail in \cite{rotor}. The position of the peaks is 
\begin{equation}
x_{c}=\pm \arccos \left( \frac{1}{\tau }\right) \mp \sqrt{\tau ^{2}-1}%
\,\,\,\,(\mbox{mod}\;2\pi ).
\end{equation}
The two peaks move in the opposite directions, and asymptotically, they have
a constant velocity $v=1$, at $\tau \rightarrow \infty .$

A sharp focal spot, and singular-like rainbow peaks are observed only at
zero initial temperature. Figures 1 (c) and (d) show the effect of the
finite temperature on the dynamics an ensemble of atoms. They present the
spatial distribution at focal time $\tau _{f}$ and at $\tau =1.84\tau _{f}$,
respectively. Instead of the singularities seen in Figs. 1 (a) and (b), we
observe broader spatial structures that are still reminiscent of the
focusing phenomenon and the rainbow effect.

In order to characterize the degree of atomic localization (squeezing), we
introduce the localization factor $L(\tau )=1-<\cos x(\tau )>$, which can be
written as 
\begin{eqnarray}
L(\tau )&=&1-\int_{-\infty }^{\infty }dv_{0}\int_{-\pi }^{\pi
}dx_{0}f(x_{0},v_{0},\tau =0) \nonumber \\
&\times& \cos \left( x_{0}-\tau \sin x_{0}+v_{0}\tau
\right) .  \label{locfac1}
\end{eqnarray}
For a thermal initial distribution, Eq.(\ref{locfac1}) can be rewritten as 
\begin{eqnarray}
L(\tau )&=&1-\frac{1}{2\pi }\int_{-\infty }^{\infty }dv_{0}f(v_{0})\cos
(v_{0}\tau ) \nonumber \\
&\times& \int_{-\pi }^{\pi }dx_{0}\cos \left( x_{0}-\tau \sin
x_{0}\right) ,
\end{eqnarray}
where $f(v_{0})=1/\sqrt{2\pi \sigma _{cl}^{2}}\exp \left[ -v_{0}^{2}/(2%
\sigma _{cl}^{2})\right] $ with $\sigma _{cl}$ defined in Eq. (\ref{sigma}).
Then, 
\begin{equation}
L(\tau )=1-\exp \left[ -\frac{\sigma _{cl}^{2}\tau ^{2}}{2}\right]
J_{1}(\tau ),  \label{cllfac}
\end{equation}
where $J_{1}(\tau )$ is Bessel function of first order. Figure 2 shows the
localization factor $L(\tau )$ for different temperatures. At zero
temperature (solid line), the localization factor takes the value of $L(\tau
_{f})\approx 0.57$ at the focal time $\tau _{f}=1$. Despite the small size
of the focal spot, the overall localization is not very marked. This is due
to a large fraction of atoms that are out of the focal area. The best
localization that can be achieved with a single kick, $L(\tau )\approx 0.42$%
, occurs after the focal event, at $\tau \approx 1.84$. The spatial
distribution at the time of the maximal squeezing for zero temperature can
be seen in Fig. 1 (b). For higher initial temperature (dashed and
dash-dotted lines in Fig. 2), the best squeezing provided by a single $%
\delta -$ pulse gets less notable, and it happens at earlier times.
\begin{figure}[b]
  \center{\epsfig{file=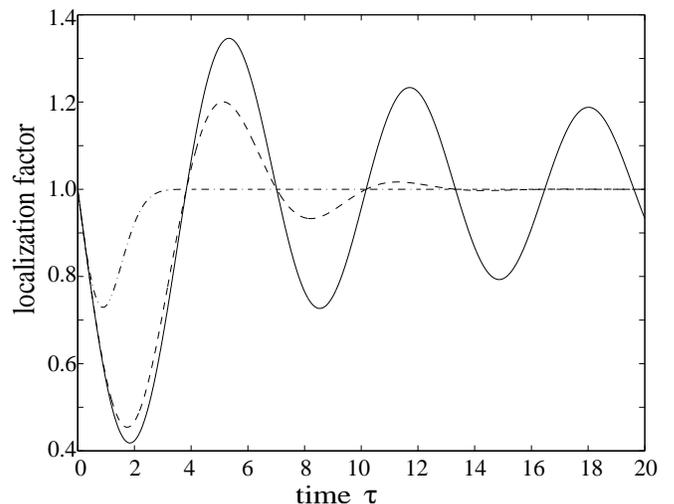,width=8.6 cm, angle=0}}
  \caption{Localization factor of a classical ensemble at different
temperatures. The solid line displays the localization factor at zero
temperature. The dashed and dash-dotted lines correspond to $\sigma _{cl}=0.2
$ and $\sigma _{cl}=1$, respectively.}
\end{figure}

\begin{figure}[t]
  \center{\epsfig{file=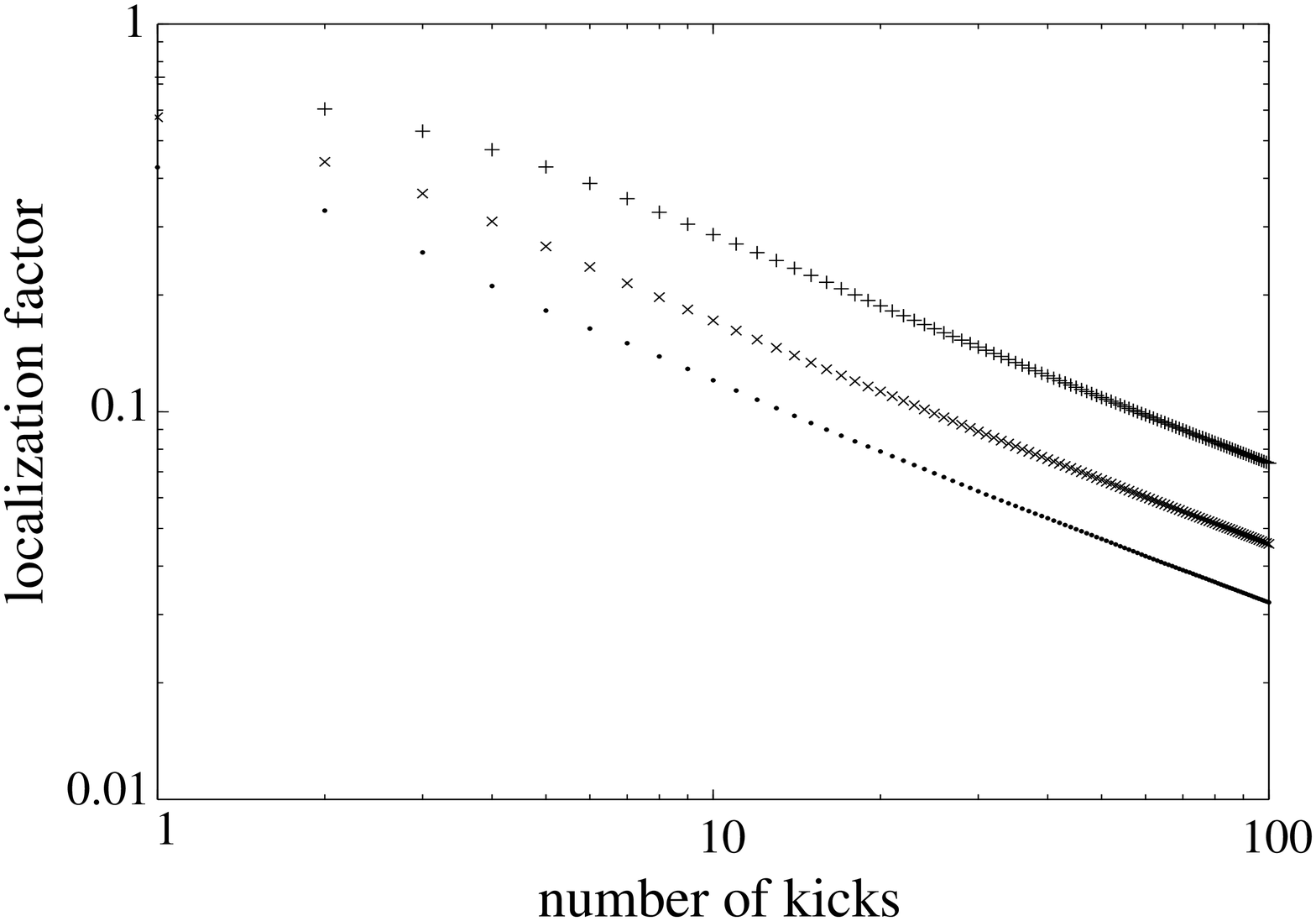,width=8.6 cm, angle=0}}
  \caption{Accumulative squeezing. The figure shows the minimal value of the
localization factor as a function of the number of kicks. The points (.)
correspond to zero initial temperature, the crosses (x) and (+) correspond
to $\sigma_{cl}=0.5$ and $\sigma_{cl} = 1$, respectively. The graphs are
shown in double logarithmic scale.}
\end{figure}

\subsection{Accumulative squeezing}

The squeezing can be enhanced by applying multiple short pulses to the
atomic system. Recently, the ''accumulative squeezing strategy'' for
efficient orientation (or alignment) of a kicked rotor was introduced in 
\cite{rotor}. In the following, we show that this strategy can be applied to
the system of cold atoms driven by a pulsed optical lattice as well. As
mentioned above, the spatial distribution of atoms is maximally squeezed at
some time $\Delta \tau _{1}$ after the first kick. According to the
''accumulative squeezing strategy'', the second kick is applied exactly at
that time, $\tau _{2}=\Delta \tau _{1}$. Immediately after the second kick,
the system has the same spatial distribution as before the kick, but $\tau
=\tau _{2}$ is no longer a stationary point for the localization factor $%
L(\tau ).$ As a result, $L(\tau )$ will reach a new minimal value at some
time point $\tau _{2}+\Delta \tau _{2}$, and this new minimum will be
smaller than the previous one. By continuing this way, we apply a sequence
of kicks at time instants $\tau _{k+1}=\tau _{k}+\Delta \tau _{k}$, and the
squeezing effect accumulates with time. We find the interval $\Delta \tau
_{k}$ between the $k$th and $(k+1)$th kicks by minimizing the localization
factor 
\begin{equation}
L(\tau _{k+1})=1-\frac{1}{2\pi }\int_{-\infty }^{\infty
}dv_{0}f(v_{0})\int_{-\pi }^{\pi }dx_{0}\cos x_{k+1},  \label{locfacacc}
\end{equation}
with respect to $\Delta \tau _{k}$\ . Here $x_{k+1}$ is defined by 
\begin{equation}
x_{k+1}=x_{k}+\Delta \tau _{k}\left( v_{0}-\sum_{i=1}^{k}\sin x_{i}\right) .
\end{equation}
Figure 3 displays $L(\tau _{k})$ for a series of 100 pulses applied at
different initial temperature. It can be seen, that after the first few
kicks the localization factor demonstrates a negative power dependence as a
function of the kick number (a straight line in the double logarithmic
scale). The squeezing is reduced for higher initial temperature. However,
the slope of all of the curves in Fig. 3 is the same after the first several
kicks, in full agreement with the asymptotics $L(\tau _{k})\thicksim 1/\sqrt{%
k}$ found in \cite{rotor}. Therefore, the accumulative squeezing scenario
may be an effective and regular strategy for atomic localization even at
finite temperature. Although finite temperature worsens the squeezing, the
temperature effects become less important with increasing the number of
kicks. The system gains kinetic energy from each pulse, so eventually the
initial thermal energy becomes small compared to the energy supplied by the
pulsed optical lattice.

\subsection{Optimal squeezing with few short pulses}

We have demonstrated that the strategy of accumulative squeezing allows, in
principle, an arbitrarily good localization of atoms. However, this does not
mean that it is the only one (or the most effective) squeezing strategy. We
have studied the best localization that can be achieved with a{\em \ given
number, }$N${\em \ } of identical $\delta $-kicks, by minimizing the
localization factor Eq.(\ref{locfacacc}) in the $N$ -dimensional space of
all possible delay times $\Delta \tau _{k}>0$. For clarity, we present here
only the results for zero initial atomic temperature. Table I shows the best
values of the localization factor found for up to five kicks, and compares
them with the results of the accumulative squeezing strategy with the same
number of pulses. While the maximal atomic localization for two pulses is
almost the same for the accumulative squeezing and for the optimal sequence
of two pulses, the optimized results for three and more pulses are much
better. The optimized delay times $\Delta \tau _{k}$ between the pulses are
shown in Table II. For three and more pulses they differ considerably from
the times between kicks according to the accumulative squeezing scheme.

For illustration, we choose the sequence of four optimal pulses to visualize
the dynamics behind the squeezing process. First, we note that the second
pulse is not applied at the time of the maximal localization. On the
contrary, the optimized procedure finds it favorable to wait a bit after the
moment of maximal squeezing, until the distribution becomes rather broad.
\begin{table}[tbp]
\begin{tabular}[t]{|c|c|c|}
\hline
No of kicks & $L_{acc}$ & $L_{op}$ \\ \hline
2 & 0.33 & 0.31 \\ 
3 & 0.26 & 0.20 \\ 
4 & 0.21 & 0.11 \\ 
5 & 0.18 & 0.07 \\ \hline
\end{tabular}
\caption{Minimal value of the localization factor that can be achieved with
a fixed number of kicks using accumulative squeezing scenario ($L_{acc}$)
and optimized sequence of pulses ($L_{opt}$).}
\end{table}
\begin{figure}[h]
  \center{\epsfig{file=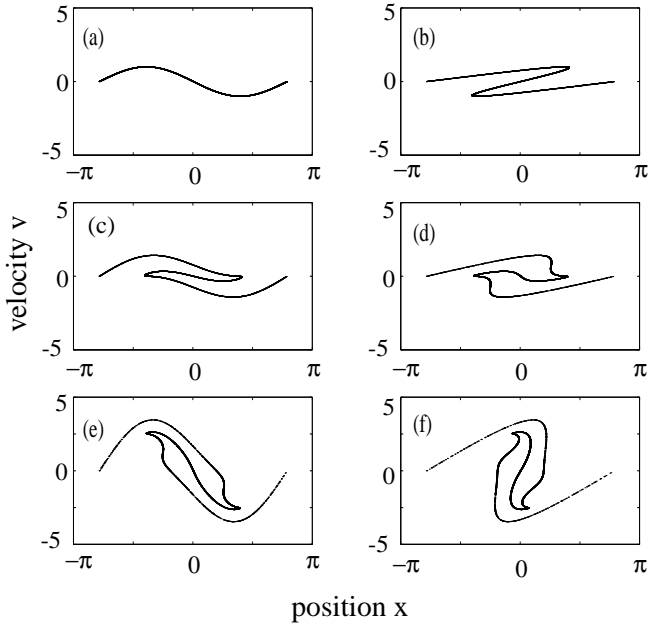,width=8.6 cm, angle=0}}
  \caption{Distribution of particles as a function of their position $x$ and
velocity $v$ for a sequence of four optimal pulses. Figs. (a) and (b) are
plotted at the time of the first pulse, and before the second pulse,
respectively. The distribution after the second pulse, and immediately
before the combined third and fourth pulses can be seen in (c) and (d).
Finally, the plots (e) and (f) show the distribution immediately after the
last pulse and at the time of maximal squeezing. Here, the initial
temperature of the ensemble is zero.}
\end{figure}
The optimal four-pulse sequence we found requires applying the third and the
forth pulses simultaneously, thus producing an effective ''double pulse''.
To understand the mechanism of the optimal squeezing with four pulses, we
refer to Fig. 4 that shows the velocity of the atoms as a function of their
position. The left column displays the velocity immediately {\em after} the
first (a), the second (c), and combined third and fourth (e) pulses. The
right column shows the velocity after the free evolution between the pulses:
immediately {\em before} the second (b), the third (d) pulse, and at time of
the final maximal localization (f). We can see that the first two pulses
fold the distribution in a way that most of the atoms are concentrated
in the harmonic region around $x=0$ (see Fig. 4 (e)) with a roughly linear
dependence of the velocity on the position.
\begin{table}[tbp]
\begin{tabular}[t]{|c|c|c|c|c|}
\hline
& $\Delta \tau _{1}$ & $\Delta \tau_{2}$ & $\Delta \tau_{3}$ & $\Delta
\tau_{4}$ \\ \hline
Accumulative squeezing & 1.84 & 0.59 & 0.42 & 0.29 \\ \hline
Optimal squeezing &  &  &  &  \\ 
2 kicks & 1.41 &  &  &  \\ 
3 kicks & 2.73 & 0 &  &  \\ 
4 kicks & 3.02 & 1.35 & 0 &  \\ 
5 kicks & 3.09 & 1.47 & 0.12 & 0.03 \\ \hline
\end{tabular}
\caption{ Comparison between the delay times $\Delta \tau_1$ - $\Delta \tau_4
$ for accumulative squeezing and optimal squeezing.}
\end{table}
Due to the double strength of
the last pulse, it emphasizes the linear character of the velocity
distribution. During the free motion of the atoms after the last pulse, the
linear-like distribution ''rotates'', so that it becomes concentrated around 
$x=0$ at the time of the maximal squeezing (see Fig. 4 (f)).

\section{Quantum squeezing}

In order to provide a quantum mechanical description for atoms in a pulsed
optical lattice, it is convenient to rescale the variables and the
Hamiltonian Eq.(\ref{Hamiltonian}) in a different way. In the classical
limit, there is a single time scale that determines the dynamics of the
kicked atoms at zero temperature, namely, the focusing time that depends on
the strength of the interaction $V(t)$. However, a quantized space-periodic
atomic motion in an optical lattice is characterized by an additional time
scale, $\tau _{r}=h/E_{r}$ where $E_{r}=\hbar ^{2}(2k_{l})^{2}/2m$ is a
kinetic energy of a particle with momentum $2\hbar k_{l}$. This time scale
determines the free evolution of an arbitrary atomic state having a spatial
period of $\lambda _{l}/2$. Although the energy spectrum of this system is
not equidistant, the quantum dynamics of free atoms repeats itself after
each period of $4\pi \tau _{r}$. This revival time, which is independent of
the interaction strength, provides a criterion for the validity of the
classical description of the atom-lattice interaction. If the relevant times
are much smaller than $4\pi \tau _{r}$, the system can be considered
classically, while for larger times quantum effects become important. We,
therefore, introduce the dimensionless time as $\tau =t/\tau _{r}$.
Furthermore, we define the dimensionless coordinate as $x^{\prime }=2k_{l}x$%
, the momentum as $p^{\prime }=p/(2 \hbar k_{l})$ and the Hamiltonian as $%
H^{\prime }=m/(2 \hbar k_{l})^{2}H$. The Hamiltonian Eq.(\ref{Hamiltonian})
can then be rewritten as 
\begin{equation}
H(x^{\prime },p^{\prime },\tau )=\frac{p^{\prime }{}^{2}}{2}-V^{\prime
}(\tau )\cos x^{\prime }.  \label{ham3}
\end{equation}
The dimensionless field intensity is $V^{\prime }(\tau )=m/(2 \hbar
k_{l})^{2} V(\tau )$. In what follows, we will drop again \ the primes for
the clarity of presentation.

If the initial atomic state is a spatially uniform one, or any other state
having the same spatial period as the optical lattice, this periodicity
property will be preserved in the course of interaction. The wave function
of the system can be the expanded in a Fourier series 
\begin{equation}
\Psi (x,\tau )=\frac{1}{\sqrt{2\pi }}\sum_{n=-\infty }^{+\infty }c_{n}(\tau
)e^{inx}.
\end{equation}
In the absence of the field, the time-dependent wave function takes the form 
\begin{equation}
\Psi (x,\tau )=\frac{1}{\sqrt{2\pi }}\sum_{n=-\infty }^{+\infty
}c_{n}(0)e^{-in^{2}\tau /2+inx}.  \label{wavefun}
\end{equation}
\begin{figure}[t]
  \center{\epsfig{file=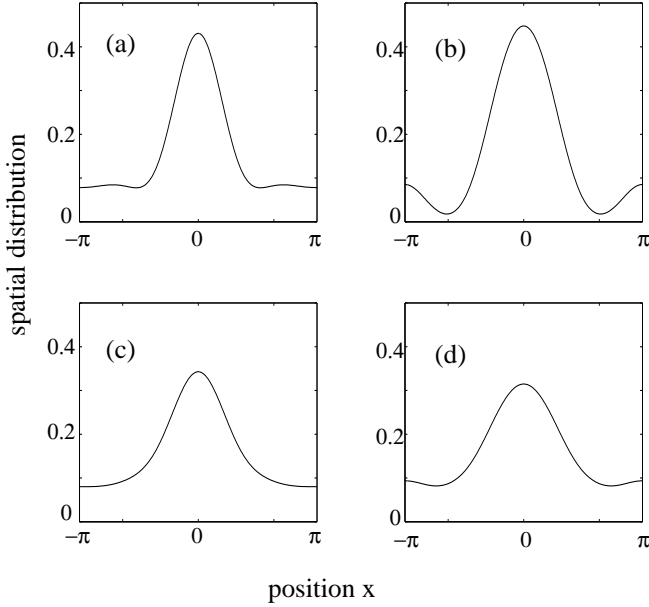,width=8.6 cm, angle=0}}
  \caption{Probability density at the classical focal time (left column) and at
the time of maximal squeezing (right column) for $P=1$. In figures (a) and
(b), the initial temperature $T$ is zero, while (c) and (d) correspond to $%
\sigma _{qm}=0.5$.}
\end{figure}
Despite the simple form of Eq.(\ref{wavefun}), the wave function exhibits an
extremely rich space-time dynamics. As mentioned before, it shows a periodic
time behavior with the period $T_{rev}=4\pi $ (full revival), and a number
of fractional revivals at $\tau =p/sT_{rev}$ ($p$ and $s$ are mutually prime
numbers) \cite{fr}. Hamiltonian (\ref{ham3}) is formally equivalent to that
one of a 2D driven quantum rotor (if the initial atomic state has a proper
space periodicity). An analytical solution valid for a general
time-dependent field is not known even for this simplest model. Much effort
has been devoted to the excitation by extremely short pulses ($\delta -$%
kicks) (see, e.g. \cite{Haake}, and references therein). In general, as a
result of a single $\delta -$kick applied to the system at $\tau =\tau _{k},$
the coefficients $c_{n}$ transform as 
\begin{equation}
c_{n}(\tau _{k}+0)=\sum\limits_{m=-\infty }^{+\infty
}i^{n-m}J_{n-m}(P)c_{m}(\tau _{k}-0),  \label{trans}
\end{equation}
where $P=\int_{-\infty }^{+\infty }V^{\prime }(\tau )d\tau $, and $J_{n}(P)$
is the Bessel function of the $n$th order. The above formulas describe only
the evolution of the states with an integer momentum. However, at finite
temperature, the initial atomic momentum has a continuous set of values.
Taking this into account, we first consider a single, non-integer initial
momentum $p_{0}=n_{0}+\nu _{0},$ where $n_{0}$ is the integer part of the
dimensionless momentum and $\nu _{0}$ belongs to the interval $[-1/2,1/2)$
(see \cite{bha99}). Due to the periodicity of the potential, the non-integer
part of the initial momentum is preserved during the interaction, so that
Eq.(\ref{wavefun}) may be replaced by 
\begin{equation}
\Psi (x,\tau )=\frac{1}{\sqrt{2\pi }}\exp (i\nu _{0}x)\sum_{n=-\infty
}^{+\infty }c_{n}(0)e^{-in^{2}\tau /2+inx}.  \label{wavefun1}
\end{equation}
\begin{figure}[t]
  \center{\epsfig{file=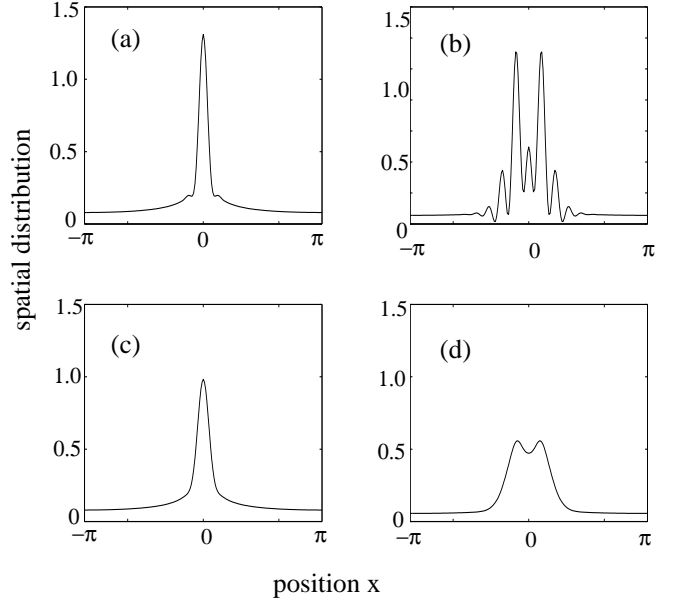,width=8.6 cm, angle=0}}
  \caption{Probability density at the classical focal time (left column) and at
the time of maximal squeezing (right column) for $P=10$. In figures (a) and
(b), the initial temperature $T$ is zero, while (c) and (d) correspond to $%
\sigma _{qm}=1$.}
\end{figure}
At finite temperature $T$, the initial distribution of momentum is 
\begin{equation}
\rho (p_{0})=\frac{1}{\sqrt{2\pi }\sigma _{qm}}\exp \left( -\frac{p_{0}^{2}}{%
2\sigma _{qm}^{2}}\right) ,  \label{thermal}
\end{equation}
where $\sigma _{qm}=\sqrt{k_{B}T m}/(2 \hbar k_l)$, with $k_{B}$ being the
Boltzmann constant. The averaged spatial probability distribution at time $%
\tau $ after a kick is then 
\begin{eqnarray}
& &\int_{-\infty }^{\infty }dp_{0}\rho (p_{0})|\Psi (x,\tau
)|^{2} \nonumber \\
&=&\sum_{n_{0}=-\infty }^{\infty }\int_{-1/2}^{+1/2}d\nu _{0}\rho
(n_{0}+\nu _{0})|\Psi (x,\tau )|^{2}.
\end{eqnarray}
Figures 5 and 6 show the spatial density of atoms after a single $\delta $
-kick, for two different pulse strengths, and different initial temperature.
The left columns show the probability density at the classical focal time ($%
\tau =1/P$), while the right columns display the density at the time of the
maximal squeezing. In Fig. 5, the pulse is rather weak ($P=1$), and the
quantum mechanical spatial distribution differs considerably from the
classical one (compare with Fig. 1). Instead of a sharp focusing spot, a
rather broad distribution centered around $x=0$ is observed even at zero
temperature (Fig. 5(a)). The finite value of the initial temperature (see
Fig. 5 (c) and (d)) broadens the distribution even more. However, for
stronger pulses ($P=10$) the quantum-mechanical results (see Fig. 6)
resemble the classical ones. At zero temperature, the quantum effects
replace the classical singularities by sharp maxima in the probability
distribution with the Airy-like shape typical to the rainbow-type scattering
in 
\begin{figure}[t]
  \center{\epsfig{file=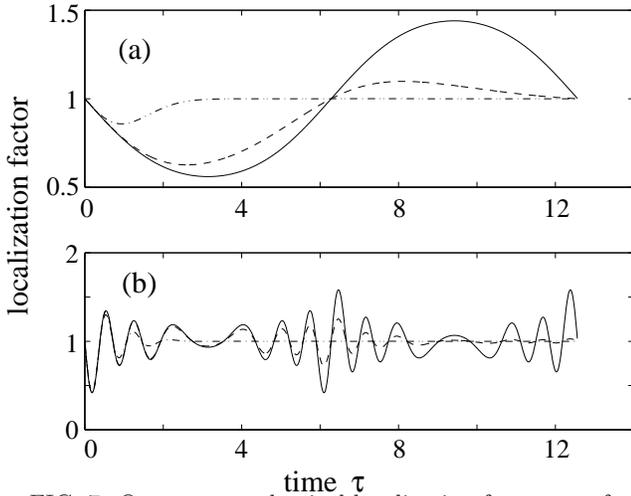,width=8.6 cm, angle=0}}
  \caption{Quantum mechanical localization factor as a function of time. The
kick strength is $P=0.5$ and $P=10$ in figures (a) and (b), respectively.
The solid lines represent the localization factor at zero initial
temperature, the dashed and dash-dotted lines correspond to $\sigma _{qm}=0.2
$ and $\sigma _{qm}=1$, respectively.}
\end{figure}
optics and quantum mechanics \cite{rotor} (see Fig. 6 (b)). At finite
temperature, however, the fine structure is washed out, and the system
behaves mainly like a classical ensemble of atoms at finite temperature
(compare Figs. 6 (c) and (d) and Fig. 1 (c) and (d)).

We proceed to the calculation of the quantum localization factor $L(\tau )=
1-<\cos x>$ for an atomic system subject to a series of short pulses applied
at $\tau =\tau _{k}$ ($k=1,2,3,...$). If the initial state contains only a
single momentum $p_{0}=n_{0}+\nu _{0},$ the localization factor after the $k$%
th pulse is 
\begin{eqnarray}
L_{p_{0}}(\tau )&=&1-Re\sum_{n=-\infty }^{\infty }c_{n}(\tau
_{k}+0)c_{n+1}^{\ast }(\tau _{k}+0) \nonumber \\
&\times&\exp \left[ i\left( n+\nu _{0}+\frac{1}{2}%
\right) (\tau -\tau _{k})\right] .
\end{eqnarray}
For a single pulse applied at $\tau =0$, one finds (using Eq. (\ref{trans})
): 
\begin{eqnarray}
L_{p_{0}}(\tau ) &=&1-\sum_{n=-\infty }^{\infty }J_{n}(P)J_{n+1}(P)\sin 
\left[ \left( n+p_{0}+\frac{1}{2}\right) \tau \right]  \nonumber \\
&=&1-\cos \left( p_{0}\tau \right) J_{1}\left( 2P\sin \frac{\tau }{2}\right)
.
\end{eqnarray}
If the system is initially in a thermal state, the localization factor has
to be averaged over the momentum distribution Eq. (\ref{thermal}) , 
\begin{equation}
L(\tau )=\int_{-\infty }^{\infty }dp_{0}\,\rho (p_{0})L_{p_{0}}(\tau ).
\end{equation}
After the averaging, it and takes the form 
\begin{equation}
L(\tau )=1-\exp \left( -\frac{\tau ^{2}\sigma _{qm}^{2}}{2}\right)
J_{1}\left( 2P\sin \frac{\tau }{2}\right) .  \label{qmlfav}
\end{equation}
Fig. 7 shows the localization factor as a function of time for various
temperatures and kick strengths, $P$. At zero temperature (solid line), the
quantum mechanical localization factor shows a revival structure, in
contrast to the classical one. The pattern, and, therefore, the state of the
maximal localization, repeats itself after each period of $4\pi $. However,
due to the continuous character of atomic momentum at finite temperature,
the revival structure disappears with increasing the temperature (see dashed
and dash-dotted lines in Figs. 7 (a) \ and (b)). In Fig. 7 (b), the strength
of the kick is $P=10$. Comparing the quantum mechanical localization factor
with its classical counterpart (see Fig. 2), we see that they have the same
structure for short times, and the same minimal localization factor $L(\tau
)\approx 0.42.$ In the short-time limit, $\tau /2\ll 1$, Eq. (\ref{qmlfav})
can be approximated by 
\begin{equation}
L(t)\approx 1-\exp \left( -\frac{\sigma _{qm}^{2}\tau ^{2}}{2}\right)
J_{1}\left( P\tau \right) ,
\end{equation}
which is exactly the expression for the classical localization factor Eq. (%
\ref{cllfac}) (we remind about the different time units used in the
classical and quantum cases). For $P=10$, maximal squeezing occurs at $\tau
=0.18$, and therefore the classical approximation is valid. For smaller
values of $P$ (for example $P=0.5$ in Fig. 7 (a)), the quantum mechanical
localization factor differs considerably from the classical one. Maximal
squeezing occurs only at $\tau \approx 3.1,$ which is well outside the
classical region. Moreover, the minimal value of the localization factor is
increasing for smaller values of $P$ (for example, $L(\tau )\approx 0.56$
for $P=0.5$). With a single kick, the most profound squeezing can,
therefore, be achieved with a relatively strong kick, after which the system
behaves more or less classically.
\begin{figure}[b]
  \center{\epsfig{file=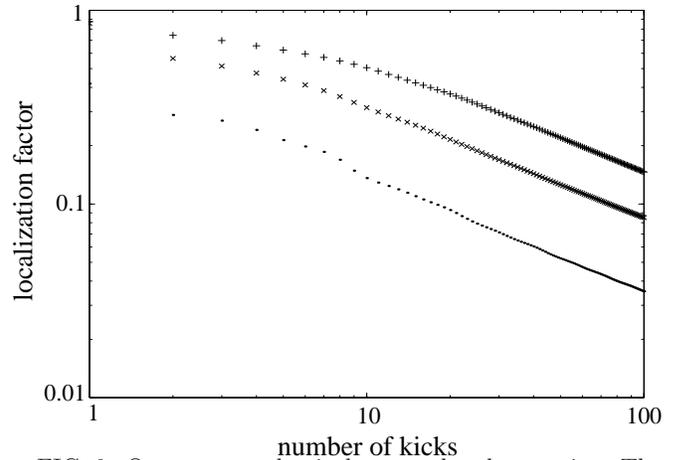,width=8.6 cm, angle=0}}
  \caption{Quantum mechanical accumulated squeezing. The figure shows the
minimal localization factor (maximal squeezing) as a function of the number
of kicks. The kick strength is $P=1$. The points (.) represent the
localization factor for zero initial temperature, and the crosses (x) and
(+) correspond to $\sigma _{qm}=1$ and $\sigma _{qm}=2$, respectively. The
graphs are shown in double logarithmic scale.}
\end{figure}
\begin{figure}[t]
  \center{\epsfig{file=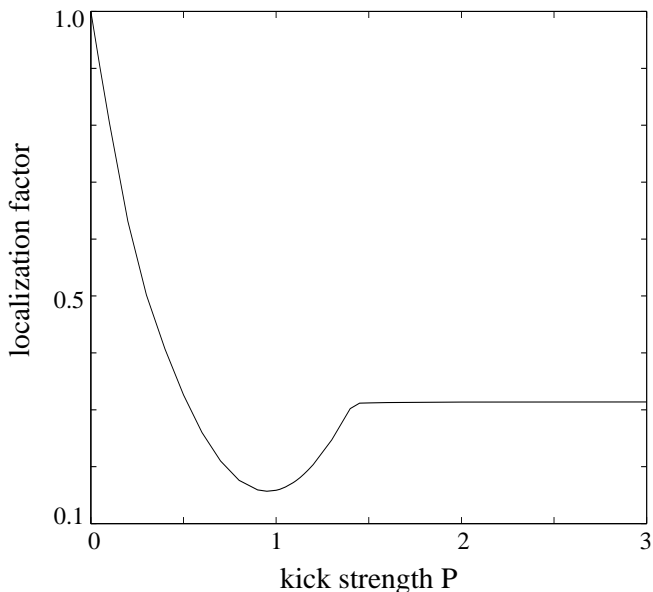,width=8.6 cm, angle=0}}
  \caption{Minimal localization factor as a function of the kick strength $P$.
Here, the optimal sequence of two identical pulses and the resulting minimal
localization factor were calculated for different values of $P$. The initial
temperature of the atoms is zero.}
\end{figure}
\subsection{Accumulative squeezing vs optimal squeezing in the quantum domain%
}

The accumulative squeezing scenario \cite{rotor} described above can be
applied to the quantized atomic motion as well. Moreover, due to the revival
phenomenon, the sequence of pulses can be made ''almost periodic'', $\tau
_{k+1}=\tau _{k}+\Delta \tau _{k}+T_{rev}$, where $\Delta \tau _{k}$ is
chosen to minimize the atomic localization factor after the $k$th kick. The
introduction of the $T_{rev}$-shift between pulses may be useful in
practical realizations of the scenario to avoid the overlap between short
pulses of a finite duration. Figure 8 demonstrates the validity of the
accumulative squeezing scenario in the quantum mechanical regime. For cesium
atoms, under conditions of a recent experiment by M. Raizen group \cite
{osk00}, the values of $\sigma _{qm}=1$ and $\sigma _{qm}=2$ correspond to
the initial temperature of $0.8 \mu$K and $3.2 \mu$K, respectively. The
system demonstrates a reduced squeezing for higher initial temperatures,
however, the slope of all of the curves in Fig. 8 becomes the same after the
first several pulses, exactly as in the classical case.

In Section II we discussed the optimal sequence of $\delta -$pulses needed
to accomplish the maximal squeezing for a given number of kicks. In the
classical limit, and at zero temperature condition, the only optimization
parameters are the time intervals (measured in the units of the focusing
time $\tau _{f})$ between the pulses. However, in the fully quantum
mechanical description, the kick strength $P$ is an independent additional
parameter. Figure 9 shows the minimal localization factor, that can be
gained with two identical $\delta -$ pulses as a function of $P$. 
\begin{figure}[h]
  \center{\epsfig{file=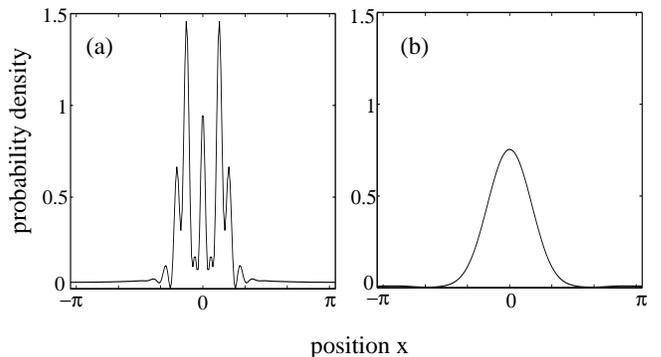,width=8.6 cm, angle=0}}
  \caption{Quantum mechanical optimal squeezing with two kicks. The figures
show the probability density of the maximally squeezed state for $P=10$ (a)
and $P=1$ (b), respectively. Here, the initial temperature of the atoms is
zero.}
\end{figure}
For $P\geq
1.5$, the best possible localization factor is almost independent of $P$.
The same value, $L\approx 0.31$, was obtained classically (see Table I). The
optimal delay time between the pulses is also in a good agreement with the
classical result $P\Delta \tau _{1}=1.4$ (see Table II). However, as can be
seen from Fig. 9, a sharp transition occurs at $P\approx 1.5$. For smaller
kick strength, the best ''quantum'' localization factor depends strongly on $%
P$, and the ''classical'' sequence of pulses is no longer the optimal
solution. The localization factor has a minimum at $P\approx 1$, and its
value, $L\approx 0.16$, is approximately half of the expected ''classical''
value. The time delay between the pulses, $\Delta \tau _{1}\approx 2.69$, is
substantially longer compared to the classically calculated interval between
the optimal kicks. Figure 10 shows the probability distribution of the
maximally squeezed atomic state after the optimal sequence of two pulses. In
Fig. 10 (a), the quantum optimization was done for $P=10$. The distribution
has a bimodal structure with two typical Airy-like rainbow peaks (compare
with the classical rainbow singularities in Fig. 1 (b)). The probability
distribution for $P=1$ (see Fig. 10 (b)) is, however, quite different from
the classical one. \ The width of the central part of the distribution shown
in Fig. 10 (b) looks comparable with the width of the distribution in Figs.
10 (a), however the localization factor in considerably smaller due to the
suppressed background.

\section{Summary}

In this paper we presented a theoretical study on the process of a transient
squeezing of an ensemble of cold atoms via interaction with a pulsed
nonresonant optical lattice. We studied in detail the atomic evolution
following a single short pulse, and described a number of semiclassical
catastrophes in the time-dependent spatial distribution of atoms, like
focusing, and rainbow formation. The problem was treated both classically,
and quantum-mechanically, and a clear correspondence between these two
descriptions was established. We have demonstrated that using a proper
sequence of short laser pulses, it is possible to narrow dramatically the
atomic spatial distribution near the minima of the light-induced potential.
In particular, we tested the strategy of accumulative squeezing that was
suggested recently \cite{rotor} for the enhanced orientation (alignment) of
a quantum rotor (molecule) interacting with short laser pulses. We showed
that this approach works effectively for the atomic squeezing, both in zero
temperature limit, and at finite temperature. Furthermore, we searched for
the {\em optimal} squeezing strategies that provide the best possible atomic
localization with a {\it given number} of short laser pulses. Both
classical, and quantum-mechanical optimization was performed. For large
enough pulse intensity, the optimal strategy for the best localization
follows the classical scenario. For smaller pulse strength, quantum
mechanical effects give rise to new localization mechanisms which may be
more effective than the ''classical'' ones. A detailed study of the optimal
squeezing approaches, also at finite temperature, will be published elsewhere.

We are glad to mention that a part of our predictions and conclusions is already 
supported by a recent
 experiment \cite{experiment} by the group of M. Raizen. 
Using the accumulative squeezing approach, they demonstrated a substantial
squeezing of cold cesium atoms kicked by a pulsed optical lattice.

\section{Acknowledgment}

We are pleased to acknowledge fruitful discussions with R. Arvieu, Y. Prior, 
and M. Raizen. The work of ML was supported in part by the Minerva Foundation.


\begin{references}
\bibitem{Jessen}  P. S. Jessen and I. H. Deutsch, Adv. Atm. Mol. Opt. Phys. 
{\bf 37}, 95 (1996).

\bibitem{Dahan}  M. BenDahan, E. Peik, J. Reichel, Y. Castin, and C. Salomon,
Phys. Rev. Lett. {\bf 76}, 4508 (1996).

\bibitem{Wilkinson}  S. R. Wilkinson, C. F. Bharucha, K. W. Madison, Qian
Niu, and M. G. Raizen, Phys. Rev. Lett. {\bf 76}, 4512 (1996).

\bibitem{Niu}  Qian Niu, Xian-Geng Zhao, G. A. Georgakis, and M. G. Raizen,
Phys. Rev. Lett. {\bf 76}, 4504 (1996).

\bibitem{Graham}  R. Graham, M. Schlautmann, and P. Zoller, Phys. Rev. A
{\bf 45}, R19 (1992).

\bibitem{Moore}  F. L. Moore, J. C. Robinson, C. F. Bharucha, B. Sundaram,
and M. G. Raizen, Phys. Rev. Lett. {\bf 75}, 4598 (1995).

\bibitem{Ammann}  H. Ammann, R. Gray, I. Shvarchuck, and N. Christensen,
Phys. Rev. Lett. {\bf 80}, 4111 (1998).

\bibitem{Timp}  G. Timp, R. E. Behringer, D. M. Tennant, J. E. Cunningham,
M. Prentiss, and K. K. Berggren, Phys. Rev. Lett. {\bf 69}, 1636 (1992).

\bibitem{McClelland}  J. J. McClelland, R. E. Scholten, E. C. Palm, and R.
J. Cellota, Science {\bf 262}, 877 (1993).

\bibitem{Cellota}  R. J. Cellota and J. J. McClelland, US Patent 5,360,764
(1994).

\bibitem{Drodofsky}  U. Drodofsky, J. Stuhler, B. Brezger, T. Schulze, M.
Drewsen, T. Pfau, and J. Mlynek, Microelectronic Engineering {\bf 35}, 285
(1997).

\bibitem{Gupta}  R. Gupta, J. J. McClelland, Z. J. Jabbour, and R. J.
Cellota, Appl. Phys. Lett. {\bf 67, }1378 (1995).

\bibitem{review}  for a recent review of atom lithography see, e.g. a
special issue on nanomanipulation of atoms, Appl. Phys. B {\bf 70}, issue 5
(2000), edited by D. Meschede and J. Mlynek.

\bibitem{Schleich}  S. Meneghini, V. I. \ Savichev, K. A. H. van Leeuwen,
and W. P. Schleich, Appl. Phys. B {\bf 70}, 675 (2000).

\bibitem{Gorlitz}  A. G\"{o}rlitz, M. Weidem\"{u}ller, T.W. H\"{a}nsch, and
A. Hemmerich, Phys. Rev. Lett. {\bf 78}, 2096 (1997)

\bibitem{Raithel1}  G. Raithel, G. Birkl, W. D. Phillips and S. L. Rolston,
Phys. Rev. Lett. {\bf 78}, 2928 (1997).

\bibitem{Bigelow}  P. Rudy, R. Ejnisman, and N. P. Bigelow, Phys. Rev. Lett. 
{\bf 78}, 4906 (1997).

\bibitem{Monroe}  C. Monroe, Nature {\bf 388}, 719 (Aug 21, 1997).

\bibitem{Birkl}  G. Birkl, M. Gatzke, I. H. Deutsch, S. L. Rolston, and W.
D. Phillips, Phys. Rev. Lett. {\bf 75}, 2823 (1995).

\bibitem{Hansch}  M. Weidem\"{u}ller, A. Hemmerich, A. G\"{o}rlitz, T.
Esslinger, and T. W. H\"{a}nsch, Phys. Rev. Lett. 75, 4583 (1995).

\bibitem{Raithel2}  G. Raithel, G. Birkl, A. Kastberg, W. D. Phillips and S.
L. Rolston, Phys. Rev. Lett. {\bf 78}, 630 (1997).

\bibitem{Burin97} A. L. Burin, J. L. Birman, A. Bulatov, and H. Rabitz,
cond-mat/9708113 (1997).

\bibitem{Bulatov98} A. Bulatov, B. Vugmeister, A. Burin, and H. Rabitz,
Phys. Rev. A {\bf 58}, 1346 (1998).

\bibitem{Bulatov99} A. Bulatov, B. E. Vugmeister, and H. Rabitz, Phys. Rev. A
{\bf 60}, 4875 (1999).

\bibitem{Normand}  D. Normand, L. A. Lompre, and C. Cornaggia, J. Phys. {\bf %
B} 25, L497 (1992).

\bibitem{Dietrich}  P. Dietrich, D. T. Strickland, M. Laberge, and P. B.
Corkum, Phys. Rev. A {\bf 47}, 2305 (1993).

\bibitem{Friedrich}  B. Friedrich and D. Herschbach, Phys. Rev. Lett. {\bf 74%
}, 4623 (1995); J. Phys. Chem. {\bf 99}, 15686 (1995).

\bibitem{Corkum}  J. Karczmarek, J. Wright, P. Corkum, and M. Ivanov, Phys.
Rev. Lett. {\bf 82}, 3420 (1999).

\bibitem{Tamar1}  T. Seideman, J. Chem. Phys. {\bf 103}, 7887 (1995); {\em %
ibid}. {\bf 106}, 2881 (1997).

\bibitem{Tamar2}  T. Seideman, Phys. Rev. Lett. {\bf 83}, 4971 (1999).

\bibitem{Cai}  L. Cai, J. Marango. and B. Friedrich, Phys. Rev. Lett. {\bf 86%
}, 775 (2001).

\bibitem{rotor}  I. Sh. Averbukh and R. Arvieu, Phys. Rev. Lett. {\bf 87},
163601 (2001).

\bibitem{patent}  I.Sh. Averbukh, patent pending.

\bibitem{Kravtsov}  Yu. A. Kravtsov and Yu. I. Orlov, {\it ''Caustics,
Catastrophes and Wave Fields''} (Springer Series on Wave Phenomena, 15), 2nd
edition, (Springer-Verlag, Berlin, 1999).

\bibitem{Ford}  K. W. Ford and J. A. Wheeler, Ann. Phys. {\bf 7}, 259 (1959).

\bibitem{Berry}  M. V. Berry, Adv. Phys. {\bf 25}, 1 (1976).







\bibitem{bha99}  C. F. Bharucha, J. C. Robinson, F. L. Moore, Bala Sundaram,
Qian Niu, and M. G. Raizen, Phys. Rev. E {\bf 60}, 3881 (1999).

\bibitem{osk00}  W. H. Oskay, D. A. Steck, V. Milner, B. G. Klappauf, and M.
G. Raizen, Opt. Comm. {\bf 179}, 137 (2000) .

\bibitem{fr}  I. Sh. Averbukh and N. F. Perelman, Phys. Lett. A {\bf 139},
449 (1989); Sov. Phys. JETP {\bf 69}, 464 (1989).

\bibitem{Haake}  F. Haake, {\it Quantum Signatures of Chaos}, Springer
Verlag, (Berlin, 1991).

\bibitem{experiment}  M. Raizen {\em et al,} to be published
\end{references}
\end{document}